# Towards using Reinforcement Learning for Scaling and Data Replication in Cloud Systems

Arar Fahem
*LMCS Research Laboratory*
Higher Nationale School of Computer Science ESI
Algiers, Algeria
cf_arar@esi.dz

Dr. Riad Mokadem
*IRIT Research Laboratory.*
*Paul Sabatier University*
Toulouse, France
riad.mokadem@irit.fr

Pr Zegour Djamel Eddine
*LMCS Research Laboratory*
Higher Nationale School of Computer Science ESI
Algiers, Algeria
d_zegour@esi.dz

*Abstract*— Given its intuitive nature, many Cloud providers opt for threshold-based data replication to enable automatic resource scaling. However, setting thresholds effectively needs human intervention to calibrate thresholds for each metric and requires a deep knowledge of current workload trends, which can be challenging to achieve. Reinforcement learning is used in many areas related to the Cloud Computing, and it is a promising field to get automatic data replication strategies. In this work, we survey data replication strategies and data scaling based on reinforcement learning (RL).

## I. INTRODUCTION

Cloud computing addresses many challenges related to the quantity of data stored and the quality of services provided. This involves meeting objectives such as improving access performance by reducing communication costs or load balancing, improving data reliability or improving fault tolerance in case node failure. Cloud users then expect the satisfaction of service level objectives (SLOs) [2]. However, in a dynamic workload environment, performance can deteriorate, which impacts compliance with the SLA, a Service Level Contract, between the supplier and its customer.

To address these problems and challenges, many works interested in scaling problems in cloud and resource allocation. In this context, data replication appears to be a promising solution. Data Replication is a well-knowing technique for addressing availability and performance issues in cloud computing, it is based on storing multiple copies of data, known as replicas, across various sites [1]. It has been commonly used in cloud systems []….

Many Cloud providers opt for threshold-based data replication to enable automatic resource scaling, given its intuitive nature. However, setting thresholds effectively needs human intervention to calibrate thresholds for each metric and requires a deep knowledge of current workload trends, which can be challenging to achieve. Dealing with this problem, workload management already uses reinforcement learning as automation solution [28],[29],[32]. Other works proposed a data replication strategy based on reinforcement learning [24]. Reinforcement Learning -RL- is a type of machine learning that is based on learning through interaction between an agent and his complex, uncertain environment by taking actions and trying to maximize a cumulative reward.

In this paper, we discuss the integration of reinforcement learning by exploring the existing literature on data scaling in the cloud using RL techniques. Then, we discuss how Reinforcement Learning techniques can be adapted and integrated into data replication. Finally we present our future work and perspectives on proposing a data replication strategy based on Reinforcement Learning.

The rest of the paper is organized as follows: Section II introduces data replication concepts and existing classifications. In section III, we discuss the motivation of integration the Reinforcement Learning in Cloud Data Replication. In section IV, we classify the scaling and data replication works based on the RL techniques used. Finally, a synthesis and perspective of our future work are given.

## II. DATA REPLICATION CONCEPTS AND EXISTING CLASSIFICATION

### A. Data Replication Concepts

Data replication consists of placing (storing) multiple copies of a data object, called replicas, on several distinct nodes [1]. Data replication in cloud try to achieve several purposes like enhance availability by making data more available to the tenants, ameliorate access by putting data copies closer to the cloud users, It helps using more parallelism across different copies on different nodes, or a better load balancing data by spreading heavily used data across different nodes. Data replication ameliorate backup and data protection using data reliability by having backups in case some data corruption scenarios, and making systems more resilient to failures by having backups ready if a node goes down.

A simple data replication solution might be to copy data to every single node in a system. But that's not really practical in cloud environment because it uses up a lot of resource like network bandwidth and storage space. So, data replication strategies in cloud must answer many points related to the specificity of cloud environments :

- When replicate ? replicating data too soon can waste resources and slow things down. But waiting too long can also cause problems because then the benefits of replication aren't useful.

- Which data to replicate? Based on the replication strategy purpose, we define the data that will be replicated.

- Where replicate? Which node will be used as destination of the replicated data. The nodes should have enough space. Also, the link and the time between the source and destination node must meet some requirements.

- How many copies should we make? A data replication strategy must define how many copies needed to meet quality standards and conditions.

- Finally, a data replication strategy designed a cloud system shoud take into account the economic costs generated by replication such as the cost generated by replication and the profitability of the supplier.

When proposing new data replication strategies, we need to think about how they fit into some specific applications and balance conflicting goals like availability, data consistency, and performance.

### B. Data Replication Strategies in Cloud :Existing Classification

In scientific literature, several synthesis studies, and surveys [2],[3],[4] have been realized into the enumeration and categorization of the principal data replication strategies in cloud environment systems. The most purpose in these works is to classify these strategies based on different criteria's :

#### 1) The nature of the strategy

Considering the nature of the data replication strategy, we will find static strategies vs dynamic strategies. In a static replication strategy [5],[6],[7], the number of data copies is decided during the planning phase before the execution. But in a dynamic replication strategy [8],[9],[10], the number and the position of the data copies is decided after the system is already up and running.

In general, static data replication strategies are easier to implement, and the selection of a static strategy mainly depends on some important factors: user access stability and pattern, node storage capacity, and bandwidth availability. However, data access methods vary widely across Cloud systems. In addition, the workload and bandwidth are not quite the same. As a result, dynamic approaches are desirable for the dynamic aspects of cloud systems. But they introduce some inconvenience, such as the difficulty in collecting accurate execution data from all nodes in the system [2].

#### 2) The control mechanism

Taking into account the control mechanism of data replication strategies, we observe centralized replication vs decentralized replications strategies. Centralized replication [11], [12], is relying on a central node that maintains an up-to-date global view of data, it is easy to implement because the replication decision is decided by the responsible node. This provides rapid response to the environment and business changes. However, the central entity must have full knowledge of all replication parameters on all system nodes. Additionally, having primary node add a complexion about fault tolerance and high availability. Decentralized replication [13], [14] solves these issues, as there is no single point of failure in the system. In Decentralized replication strategies the system behaves as predicted even if several nodes are disconnected or faulty. However, the lack of central control or the failure to consider can result to incomplete information about the state of the system leading to issues such as over-replication.

#### 3) The orientation of the profit

By analyzing the orientation of the profit considered in the replication strategies, we uncover provider-centered strategies, that tries to maximize profitability of the cloud provider's vs tenants-centered strategies that tries to minimize cost and ameliorate service delivery [2].

Most of the replication strategies proposed in cloud infrastructure focus on reducing the resource consumption for the provider once the tenants' objectives are met. The profit of the provider is met by adjusting resource consumption like storage and CPU and reducing bandwidth consumption [14], [15], [16].

On the other hand, few data replication strategies focus on the cost that tenants pay to the provider. The profit of the tenants is met by selecting the server configuration that best fits its needs in terms of performance and cost [17],[18].

#### 4) The nature of the cloud environment

While most of the replication strategies proposed in the literature were proposed in a single-provider cloud environment, a few strategies have been deployed in an interconnected cloud environment [13].

In multi-cloud strategies, the price gap between cloud providers is leveraged to minimize costs while considering requirements for fault tolerance and latency. For other strategies [19], tenants rent services from multiple providers based on a pricing policy that considers the prices of resources provided by each cloud provider.

#### 5) The considered type and number of objectif function

Data replication strategies can be divided into two main types based on the number of objectives treated: single-objective and multi-objective strategies, each aimed at optimizing different aspects of system performance. Main objectives considered in these strategies include availability [20], which is based on the availability of the system nodes and replicated data copies, fault tolerance [21], that provides continuous availability for applications and data by creating, regenerating, and replicating data after a node failure. Energy consumption reduction [22], by optimizing resource allocation for sustainable resources and reducing energy consumption and carbon emissions.

Performance objective such as response time [13], generally is not included and guaranteed in providers Service Level Agreements (SLAs) due to the heterogeneous workloads in cloud systems. Performance objective can conflict with the goal of maximizing economic benefits while minimizing operating costs [8]. Only few replication strategies include performance objectives

An objective function, which is a parameter to optimize system performance, describes the way each system is designed to achieve a specific performance objective. Single objective strategies focus on increasing the data area, network bandwidth and environment to use, or reduce replication costs based on cost models , multi-objective strategies consider economic actions to reduce replication costs and meet other objectives simultaneously. By using these techniques, systems can efficiently manage data replication to reduce energy consumption and costs while increasing overall performance and reliability.

Many Cloud providers opt for threshold-based data replication to enable automatic resource scaling, given its intuitive nature. however, setting thresholds effectively needs human intervention to choose best thresholds for each metric and requires a global knowledge of the current workload trends, which can be challenging to achieve.

Machine learning and more specifically reinforcement learning is a promising solution in cloud data replication strategies. Many works interested in scaling problems, workload management, and resource allocation in cloud, already uses reinforcement learning as automation solution.

In the next session we explore the possibility of using reinforcement learning in cloud data replication strategies by comparing and studying the RL used in cloud scaling problems.

## III. INTEGRATION OF REINFORCEMENT LEARNING IN CLOUD DATA MANAGEMENT: MOTIVATION

### A. Reinforcement Learning

Machine learning is a subfield of artificial intelligence that focuses on the design and development of algorithms that can automatically learn from data, without being explicitly programmed. Authors in [23] define Machine learning as "A computer program is said to learn from experience E with respect to some tasks T and some performance P, if its performance on T, as measured by P, improves with experience E".

Machine learning algorithms can be classified into three main categories: supervised learning [23], unsupervised learning [23], and reinforcement learning [24]. Supervised learning algorithms are trained on labeled data, where the desired output is known in advance. Unsupervised learning algorithms work with unlabeled data and are used to identify patterns and relationships within the data. Reinforcement learning algorithms use feedback from the environment to improve their performance over time.

Reinforcement learning (RL) is a type of machine learning where an agent learns to make decisions by performing actions in an environment to maximize a reward signal. RL is defined as a way of programming agents by reward and punishment without needing to specify how the task is to be achieved [23]. It's used to solve problems where the outcome is uncertain, and the agent must learn to map situations to actions to maximize reward.

In RL, an agent interacts with an environment, taking actions and observing the resulting state and reward. The agent's goal is to learn a policy, which is a mapping from states to actions, that maximizes the expected cumulative reward over time. The learning process involves trial and error, as the agent explores the environment and updates its policy based on the received rewards [23].

Supervised learning is learning from a training set of labeled examples provided by a knowledgeable external supervisor. Each example is a description of a situation together with a specification-the label- of the correct action the system should take to that situation, which is often to identify a category to which the situation belongs. The object of this kind of learning is for the system to generalize, its responses so that it acts correctly in situations do not present in the training set. This is an important kind of learning, but alone it is not adequate for learning from interaction like our case. In interactive problems like data replication, it is often impractical to obtain examples of desired behavior that are both correct and representative of all the situations in which the agent must act. In uncharted territory-where one would expect learning to be most beneficial- an agent must be able to learn from its own experience [23].

Unsupervised learning consists in finding structure hidden in collections of unlabeled data. The terms supervised learning and unsupervised learning would seem to classify machine learning paradigms, but they do not, reinforcement learning is trying to maximize a reward signal instead of trying to find hidden structure. Uncovering structure in an agent's experience can certainly be useful in reinforcement learning, but by itself does not address the reinforcement learning problem of maximizing a reward signal [23].

Also, conventional supervised learning approaches divide the process into distinct stages, involving a strictly defined learning period based on past feedback and a subsequent stage for making predictions. The decoupling of these stages is restrictive as they are disjointed, and incorporating new data patterns into the model typically necessitates a new training period. The utilization of techniques that merge both stages, learning and prediction, such as reinforcement learning (RL), holds the promise of overcoming this limitation [23] [24].

### B. Reinforcement Learning In Cloud Scaling and data replication

Even the use of reinforcement learning (RL) has shown a big success in many fields like robotics, games, and autonomous systems, its integration into data replication strategies in the cloud environments remains relatively unexplored. Despite the absence of RL implementations in this context, there exists many works that addressing scaling issues in cloud computing using RL with different methods and ways.

Scaling in cloud is defined as adjusting the resources (such as storage, processing, and memory) allocated to the infrastructure to handle increasing amounts of data. Scaling can take two forms: (i) horizontal scaling, when the number of assigned VMs of any type to an application can dynamically vary through its execution, and (ii) vertical scaling, when the capabilities of individual VMs are varied without hindering the execution of the applications in such VMs [25]

Generally, the purpose of data replication is to enable, through data redundancy, improvement of availability and/or performance or even disaster recovery. By replicating data across multiple nodes or data centers, a data replication strategy permits to avoid the risk of data loss and ensure access to critical data, or to improve economic benefit of the cloud provider. On the other hand, data scaling is primarily geared towards enhancing system performance and scalability. It involves dynamically adjusting resources, such as adding or removing servers or upgrading existing hardware to meet changing demands and maintain optimal performance levels.

While data replication and scaling serve distinct purposes and utilize different implementation approaches, they are not mutually exclusive concepts. In fact, they are often complementary and synergistic in cloud architectures. Data Replication is a horizontal scaling [34], and both mechanisms are triggered by the same conditions and objectives.

In the next section, we seek to highlight the use of reinforcement learning in scaling problems, and the possibility of adaption the RL in data replication strategies.

## IV. EXISTING WORK IN DATA REPLICATION AND SCALING BASED ON REINFORCEMENT LEARNING

There are several techniques for obtaining adequate policies in reinforcement learning. One approach is the model-based approach, which relies on having a perfect and a complete environment model to get the appropriate policies in advance or in offline mode. In this approach Dynamic Programming is used to get the Value Iteration. Conversely, there are model-free methods, without requiring a perfect model of the environment, this method allow an agent to learn and adjust policy in real time, the policy is learned and improved over time in a process of continuous interaction with the environment [23]. In the Model-free category, we find the use of Q-learning and SARSA algorithms in temporal Difference learning.

Reinforcement learning (RL) techniques in cloud environments face challenges when dealing with extensive state spaces. These challenges directly influence the performance of such algorithms, affecting both the time required to compute a solution and the memory resources consumed. To address these issues, researchers have proposed the utilization of non-linear functions for approximating Q(s, a). Additionally, integrating RL with deep neural networks, giving rise to what is known as Deep Reinforcement Learning (DRL).

In scientific literature, few synthesis studies, and surveys [37] have been realized into studying the use of Reinforcement learning (RL) in scaling problems in the cloud systems. Our purpose is studying the possibility of integrating reinforcement learning techniques used generally in scaling problems to the data replication strategies in cloud.

Many scaling [26],[27],[28],[29],[30],[31],[32],[33] and data replication [24] works can be classified based on the reinforcement learning techniques used :

### A. Model-based Approaches

There is only one proposal [26] from our list in the Model-based category. due to the requirement of having a complete model of the environment in these methods, there is the estimation of the probability distribution of the transition between states. Also, we find a limitation of learning the policies in an offline mode while operating in a dynamic environment which is the Cloud infrastructure.

In [26], the work examines the development of budget allocation policies for workflow autoscaling in cloud environments. They utilize an MDP model constructed from the outputs of multiple workflow executions to derive appropriate policies using Value Iteration. The considered states combine the current budget limit, and the probability of out-of-bid errors, with values arrived to create 192 possible states. Actions represent different budget assignment ratios between spot and on-demand instances. Rewards are calculated based on the ratio between workflow execution progress and cost in the preceding cycle.

### B. Model-free Approaches

In contrast to Model-based approaches, Model-free methods employ an online learning approach and do not rely on having a perfect model of the environment. These methods are categorized based on their approach: sequential learning, parallel learning, and proposals combined with neural networks or Deep reinforcement learning.

*1) Sequential learning process*

In sequential learning process, each decision time, the value of a single state-action pair is updated in the table of values Q(s, a), for these reasons the process has long training times since the speed of convergence of the RL algorithms depends directly on the dimension of the state space and actions. The [27] present an autonomous solution to the problem of dynamic adaptation of the number of resources allocated to Cloud applications. The solution is based on the Q learning algorithm. For the definition of the states the number of user requests per second, the number of VMs assigned to the application, and the average response time of requests are considered. The actions represent the number of VMs to acquire or release between -1 and 10, while the reward considers the cost of acquiring or maintaining the VMs and a penalty for Service Level Agreement (SLA) violation.

The reference [30] address the automatic autoscaling of virtualized firewalls in a Cloud. The work proposes an automatic auto scale based on RL that decides when it is convenient to increase or reduce the number of active firewalls by dynamically adjusting the system to avoid overloading or wasting resources. The Q-learning algorithm is used to arrive to a scaling policy, states are defined by the current request rate and the number of active firewalls. The actions consist of increasing, reducing, or maintaining the number of active firewalls, and the reward is responsible for penalizing overload or low load states, as well as SLA violations [30].

In the reference [31], The Q-learning algorithm is used for decision-making agent to learn when to add or remove VMs to balance between SLA and costs. The states are based on the CPU utilization with three possible state or value: normal-utilization, under-utilization and over-utilization compared to a fixed threshold. The work defines three possible actions based on the three states: scale in that mean increasing, scale-out that mean decreasing and no-action that means keep or maintain the size of the infrastructure. The reward was defined by using a function R assigns a fixed value for each state-action pair. The results showed that this approach was able to reduce by 50% the total cost and increase the use of resources by 12%.

In reference [33], the proposed reinforcement learning based auto-scaler resolve the challenge of managing multi-user workloads in cloud and edge data centers by extending the single-user's workload Markov Decision Process (MDP) framework. The states represent system conditions such as CPU usage, peak latency, and the number of active Cloud-native Network Functions (CNFs) per user. The RL agent interacts with the environment to optimize resource allocation, the actions are increase, decrease, or maintain the number of CNF instances per user. The reward objective function is to maintain peak latency and CPU usage within predefined thresholds, while penalizing termination situations like very high latency or the number of CNF instances surpasses the maximum capacity of the infrastructure. The reward function is designed to treat all users equally and encourages the agent to balance resource allocation efficiently to meet Service Level Agreement (SLA) requirements. The algorithm used is PPO (Proximal Policy Optimization) [23] a specific algorithm within the Policy Gradient family, which aims to improve stability and sample efficiency, Policy Gradient family -PPO-are a class of reinforcement learning algorithms used to directly optimize the policy function to maximize expected rewards.

*2) Parallel learning process*

In parallel learning process, multiple agents that are sharing the knowledge are used to accelerate the learning and convergence time, the purpose is to get a good policy in less time, but the design is more complicated.

The reference [28] propose a method based on MDP and Q-learning for dynamic scaling in cloud. The states are defined based on the number of user requests, the number of VMs of each type, region, and the Coordinated Universal Time. Actions are either requesting, maintaining, or removing instances, while the reward includes the cost and a penalty in case of SLA violations. The work accelerates the convergence of the Q learning algorithm by using parallel learning using multiple agents.

The reference [29] propose an RL-based approach for automatically scaling virtualized resources in the Cloud. The states are defined based on the number of user requests, the infrastructure utilization based on the VMs used and acquired, also the response time and performance observed for each task during a pre-determined period. The actions are the scale-up or scale-down of the number of VMs.

*3) Deep Reinforcement Learning*

In Deep Reinforcement Learning -DRL-, the RL process is combined with Deep neural networks to resolve the problem of the state sizing. DRL can accommodate larger search spaces. This mechanism incorporates a complex neural network on top of the RL algorithms like Q-learning [24].

The authors in [32], propose a work that use RL techniques for horizontal scaling in the Cloud by adjusting resource to balance between performance and cost. States are defined by the number of VMs and various performance metrics including CPU Utilization, Network Packets, and Latency. Actions are increasing or reducing infrastructure by one or two VMs or keeping the current resources. The rewards are determined by the cost of provisioned resources and a penalty related to CPU utilization, which is contingent on the SLA agreement. The work compares three RL strategies: tabular Q-learning (QL), deep Q-network (DQN), and double-dueling Q-network (D3QN), in the CloudSim simulator and the Amazon Cloud.

The reference [24] presents a machine learning mechanism based on reinforcement learning that attaches to a hybrid replication middleware connected to a DBMS to dynamically live-tune the configuration of the middleware according to the workload being processed. The algorithm chosen was the Deep Q-learning algorithm, as it can accommodate larger search spaces. This mechanism incorporates a complex neural network on top of the Q-learning algorithm. Given this choice, the action space was sampled into a subset of 10 possible choices. As the environment being considered is bounded to database replication and overall database execution, the impact can be directly associated with the overall throughput. A higher throughput represents a better outcome. Consequently, the reward function is associated with the throughput, being defined as the sum of all latest metric averages that refer to replica throughput [24].

*C. Synthesis and Comparaison*

Most of the studied papers, in this survey, are based on model-free approaches which is justified by the complexity of the cloud environment and the requirement of having a complete model of the environment. In fact, it is not possible all the time to get the estimation of the probability distribution of the transition between states. Even we get these probabilities, they are changing all the time due to the dynamism of the cloud environment. Also, offline learning is used in model-based algorithms, which is not the most adequate for a changing environment.

The Q learning and SARSA algorithms are most used in sequential and parallel reinforcement learning model, but also DRL appears as a solution of the large size of the states, and an accelerator of convergence and learning time.

Finally, most of studied papers in this work are treating scaling problems in cloud, there is only a few works that use reinforcement learning in data replication strategies.

## V. CONCLUSION AND FUTURE WORK

While threshold-based data replication offers intuitive automatic resource scaling for many Cloud providers, setting thresholds effectively needs human intervention to adjust thresholds and requires a deep knowledge of current workload. This is a challenging task regarding the architecture, heterogeneous and complexity of cloud environment. For these reasons, machine learning and more specifically reinforcement learning constitutes a promising solution when managing resources in cloud environments.

Many works in cloud computing interested in scaling problems, resource allocation and workload management. Only few work use reinforcement learning as automation solution. Some of them were interested in NoSQL systems [35] and only a few reinforcement learning based automatic scaling works in the Cloud are dedicated to querying relational databases. Existing methods must then be adapted to the context of relational databases, considering numerous dependent tasks and intermediate relationships which can be stored on disk.

Our future work will be based on designing an efficient data replication strategy based on reinforcement learning. The proposed strategy could be oriented to the profit of the cloud provider or to reduce the cost paid by the tenant. Then, implement the strategy via simulation [36]. We will focus on relational database management systems (DBMS) and OLAP applications.

REFERENCES

[1] P. A. Bernstein, V. Hadzilacos, N. Goodman. Concurrency Control and Recovery in Database Systems. Massachusetts, Addison-Wesley Publishers, Book ISBN 0-201-10715-5, (1987).J. Clerk Maxwell, A Treatise on Electricity and Magnetism, 3rd ed., vol. 2. Oxford: Clarendon, 1892, pp.68–73.

[2] Mokadem, R., Martinez-Gil, J., Hameurlain, A. and Kueng, J. (2022) 'A review on data replication strategies in cloud systems', Int. J. Grid and Utility Computing, Vol. 13, No. 4, pp.347–362.

[3] Milani BA, Navimipour NJ. A comprehensive review of the data replication techniques in the cloud environments: Major trends and future directions. J Netw Comput Appl. 2016;64:229-238.

[4] Hamrouni, T., Mokadem, R., & Khelifa, A. (2023). Review on data replication strategies in single vs. interconnected cloud systems: Focus

on data correlation-aware strategies. Concurrency and Computation: Practice and Experience, 35.

[5] S. S. Begum and S. Sirisha. Cloud optimal security using data fragmentation and replication. Int. J. Of Computer Science Ingeneering and scientific technology (IJCSEST). ISSN 6201 3454, (2019).

[6] Z. Zeng, B. Veeravalli. Optimal metadata replications and request balancing strategy on cloud data centers. J Parallel Distrib Comput; vol. 74, Issue 10: pp. 2934–2940, (2014).

[7] S-Q. Long, Y-L. Zhao and W. Chen. MORM: A Multi-objective strategy for cloud storage cluster. Journal of Systems Architecture, Vol. 60, Issue 2, pp. 234–244, (2014).

[8] R. Mokadem, A. Hameurlain. A Data Replication Strategy with Tenant Performance and Provider Economic Profit Guarantees in Cloud Data Centers. Journal of Systems and Software (JSS), Elsevier, V. 159, https://doi.org/10.1016/j.jss.2019.110447, (2020).

[9] K. Tabet, R. Mokadem and M. R. Laouar. A Data Replication Strategy for Document Oriented NoSQL Systems. Int. Journal of Grid and Utility Computing, pp. 53-62 .(2019)

[10] N. Mansouri, N, M. M. Javidi. A hybrid data replication strategy with fuzzy-based deletion for heterogeneous cloud data centers. J Supercomput, Volume 74, Issue 10, pp. 5349–5372, (2018a)

[11] S. S. Begum and S. Sirisha. Cloud optimal security using data fragmentation and replication. Int. J. Of Computer Science Ingeneering and scientific technology (IJCSEST). ISSN 6201 3454, (2019)

[12] Y. Zhang, J Jiang, K Xu, X. Nie, M. j. Reed, H. Wang, G. Yao, , M. Zhang, K. Chen. BDS: a centralized nearoptimal overlay network for inter-datacenter data replication. The 30th EuroSys Conference, pp. 1–14, (2018)

[13] Y. Mansouri, R. Buyya. Dynamic replication and migration of data objects with hot-spot and cold-spot statuses across storage data centers. J. of Parallel and Distributed Computing. Vol. 126, pp. 121-133, (2019)

[14] U. Tos, R. Mokadem, A. Hameurlain, T. Ayav, S. Bora. Ensuring performance and provider profit through data replication in cloud systems. Cluster Computing, Vol. 21 Issue 3, pp. 1479-1492, (2018)

[15] J. Liu, H. Shen, H. S. Narman, Z. Lin, Z. Li. Popularity-aware Multi-failure Resilient and Cost-effective Replication for High Data Durability in Cloud Storage. Transactions on Parallel and Distributed Systems. (2018)

[16] U. Tos, R. Mokadem, A. Hameurlain, T. Ayav, S. Bora. A Performance & Profit Oriented Data Replication Strategy for Cloud Systems, IEEE Conf. on Cloud and Big Data Computing, CBDCom,France, pp. 780–787, (2016)

[17] S. Limam, R. Mokadem, G. Belalem. Satisfying Availability and Performance Requirements while Ensuring Profit of Cloud Providers. Int Conf. on Embedded & Distributed Systems (EDiS), Oran, Algeria,(2017).

[18] U. Sharma, P. Shenoy, S. Sahu, A. Shaikh. Kingfisher: Cost-aware elasticity in the cloud. Proceedings IEEE INFOCOM. DOI: 10.1109/INFCOM.2011.5935016, (2011)

[19] A. Abouzamazem and P. Ezhilchelvan. Efficient inter-Cloud réplication for high-availability services. Int. Conf. on Cloud Engineering (IC2E), pp. 132–139. IEEE, (2013).

[20] K. Liu, J. Peng, J. Wang, W. Liu, Z. Huang, J. Pan. Scalable and Adaptive Data Replica Placement for GeoDistributed Cloud Storages. IEEE Trans. on Parallel and Distributed Systems, vol. 31, no. 7, pp. 1575-1587, (2020).

[21] S. A. E. Selvi, R. Anbuselvi. RAAES : Reliability-Assured and Availability-Enhanced Storage for Cloud Environment. Int. Journal of Pure and Applied Mathematics, vol. 118, pp. 103–112. (2018).

[22] E. B. Edwin, P. Umamaheswari, M.R. Thanka. An efficient and improved multi-objective optimized replication management with dynamic and cost aware strategies in cloud computing data center. Cluster Computing, Vol. 22, Supp. 5, pp. 11119-11128, (2019).

[23] Sutton, R. S. (2018). Reinforcement learning : an introduction. Cambridge, MA : The MIT Press.

[24] Ferreira, L. C. (2020). Ferreira, L., Coelho, F., & Pereira, J. Self-tunable DBMS Replication with Reinforcement Learning. Distributed Applications and Interoperable Systems,. 12135, 131 - 147.

[25] Jakub Krzywda, Ahmed Ali-Eldin, Trevor E. Carlson, Per-Plov Ostberg, and Erik Elmroth. Power performance tradeoffs in data center servers: DVFS, CPU pinning, horizontal, and vertical scaling. Future Generation Computer Systems, 81:114–128, 2018.

[26] Garí, Y., Monge, D.A., Mateos, C., & García Garino, C. (2019). Learning budget assignment policies for autoscaling scientific workflows in the cloud. Cluster Computing, 23, 87-105.

[27] Dutreilh, X., Kirgizov, S., Melekhova, O., Malenfant, J., Rivierre, N., & Truck, I. (2011). Using Reinforcement Learning for Autonomic Resource Allocation in Clouds: towards a fully automated workflow.

[28] Barrett, E., Howley, E., & Duggan, J. (2013). Applying reinforcement learning towards automating resource allocation and application scalability in the cloud. Concurrency and Computation: Practice and Experience, 25.

[29] Benifa, J.V., & Dharma, D. (2018). RLPAS: Reinforcement Learning-based Proactive Auto-Scaler for Resource Provisioning in Cloud Environment. Mobile Networks and Applications, 24, 1348 - 1363.

[30] Dezhabad, N., & Sharifian, S. (2018). Learning-based dynamic scalable load-balanced firewall as a service in network function-virtualized cloud computing environments. The Journal of Supercomputing, 74, 3329-3358.

[31] Ghobaei-Arani, M., Jabbehdari, S., & Pourmina, M.A. (2018). An autonomic resource provisioning approach for service-based cloud applications: A hybrid approach. Future Gener. Comput. Syst., 78, 191-210.

[32] Wang, Z., Gwon, C., Oates, T., & Iezzi, A. (2017). Automated Cloud Provisioning on AWS using Deep Reinforcement Learning. ArXiv, abs/1709.04305.

[33] Jimenez, J., Soto, P., Vleeschauwer, D.D., Chang, C., Bock, Y.D., Latré, S., & Camelo, M. (2023). Resource Allocation of Multi-User Workloads in Cloud and Edge Data-Centers Using Reinforcement Learning. 2023 19th International Conference on Network and Service Management (CNSM), 1-5.

[34] G. Graefe, P. Alto, A. Nica, K. Stolze, T. Neumann, T. Eavis, I. Petrov, E. Pourabbas, A. Rubertiand, D. Fekete. Elasticity in Cloud Databases and Their Query Processing. International Journal of Data Warehousing and Mining (IJDWM), Vol. 9 Issue 2, pp. 1-20, (2013)

[35] A. Naskos, A. Gounaris, I. Konstantinou. Elton: a cloud resource scaling-out manager for nosql databases. 34th IEEE Int. Conf. on Data Engineering (ICDE), IEEE, pp.1641–1644. (2018)

[36] R.N. Calheiros, R. Ranjan, A. Beloglazov, C.A.F. De Rose, R. Buyya. CloudSim: A Toolkit for Modeling and Simulation of Cloud Computing Environments and Evaluation of Resource Provisioning Algorithms. Software: Practice and Experience. V. 41, N. 1, pp. 23-50. (2010)

[37] Gar'i, Y., Monge, D.A., Pacini, E., Mateos, C., & Garino, C.G. (2020). Reinforcement learning-based application Autoscaling in the Cloud: A survey. Eng. Appl. Artif. Intell., 102, 104288.